\documentclass[aps,pra,showpacs,twocolumn,amsmath,amssymb,superscriptaddress,foo
tinbib]{revtex4}

\usepackage[english]{babel}
\usepackage{times}
\usepackage{latexsym}

\usepackage{graphicx} 
\usepackage{epsf} 
\usepackage{subfigure} 
\usepackage{amsmath} 
\usepackage{amssymb} 
\usepackage{amsfonts}
\usepackage{bm}
\usepackage{natbib}
\usepackage{epstopdf}

\def\be{\begin{equation}}
\def\ee{\end{equation}}
\def\bea{\begin{eqnarray}}          
\def\eea{\end{eqnarray}}
\def\bi{\begin{itemize}}
\def\ei{\end{itemize}}
\def\bin{\begin{enumerate}}
\def\ein{\end{enumerate}}

\begin{document}

\title{Ultracold atomic gas in non-Abelian  gauge potentials: the quantum Hall effect supremacy}


\author{
N. Goldman
}
\affiliation{Center for Nonlinear Phenomena and Complex Systems - Universit$\acute{e}$ Libre de Bruxelles (U.L.B.), Code Postal 231, Campus Plaine, B-1050 Brussels, Belgium}

\author{A. Kubasiak 
}
\affiliation{
ICFO-Institut de Ci\`encies Fot\`oniques,
Parc Mediterrani de la Tecnologia,
E-08860 Castelldefels (Barcelona), Spain}

\affiliation{Marian Smoluchowski Institute of Physics Jagiellonian University, Reymonta 4, 30059 Krak\'ow, Polska
}

\author{
P. Gaspard
}
\affiliation{Center for Nonlinear Phenomena and Complex Systems - Universit$\acute{e}$ Libre de Bruxelles (U.L.B.), Code Postal 231, Campus Plaine, B-1050 Brussels, Belgium}

\author{ 
M. Lewenstein}

\affiliation{
ICFO-Institut de Ci\`encies Fot\`oniques,
Parc Mediterrani de la Tecnologia,
E-08860 Castelldefels (Barcelona), Spain}
\affiliation{
ICREA - Instituci\`o Catalana de Ricerca i Estudis Avan{\c c}ats, 08010 
Barcelona, Spain}

\date{\today}

\begin{abstract}
Nowadays it is experimentally feasible to create artificial, and in particular, non-Abelian
gauge potentials for ultracold atoms trapped in optical lattices. 
Motivated by this fact, we investigate the fundamental properties of an ultracold Fermi gas
in a  non-Abelian $U(2)$ gauge potential characterized by a \emph{constant} Wilson loop. Under this specific condition, the energy spectrum  exhibits a robust band structure with large gaps and reveals a new fractal figure. The transverse conductivity is related to topological invariants and is shown to be quantized when the Fermi energy lies inside a gap of the spectrum. We demonstrate that the analogue of the integer quantum Hall effect for neutral atoms survives the non-Abelian coupling and leads to a striking fractal phase diagram. Moreover, this coupling induces an anomalous Hall effect as observed in graphene.
\end{abstract}

\maketitle
\section{Introduction}
Ultracold atoms in optical lattices offer unprecedented possibilities of controlling quantum matter and mimicking the systems of condensed-matter and high-energy physics \cite{Bloch, ouradv}.  Particularly fascinating is the possibility to study ultracold atoms under the influence of strong artificial Abelian and non-Abelian  ``magnetic" fields. The experimental realization of artificial Abelian ``magnetic" fields, which reproduce the physics of electrons in strong magnetic fields, is currently achieved through diverse schemes: for atoms in a trap the simplest way is to rotate the trap \cite{ouradv}, while for atoms in optical lattices this can be accomplished by combining laser-assisted tunneling and lattice acceleration methods \cite{Jaksch,Mueller,Demler,Ohberg,Ohberg2,spielman}, by the means of lattice rotations \cite{Holland1,Holland2,Tung}, or by the immersion  of atoms in a 
lattice within a rotating Bose-Einstein condensate (BEC) \cite{imme}.            
Several phenomena were predicted to occur in these arrangements such as the Hofstadter ``butterfly" \cite{Hofstadter} and the ``Escher staircase" \cite{Mueller}
in single-particle spectra, vortex formation \cite{ouradv, Holland1, Goldman2}, quantum  Hall effects \cite{Demler,Palmer,Goldman1, Holland2}, as well as other 
quantum correlated liquids \cite{Hafezi}.
  
As shown by one of us in Ref. \cite{Osterloh}, it is simple to generalize the scheme of Jaksch and Zoller 
 for generating artificial Abelian ``magnetic" fields \cite{Jaksch} in order to mimic artificial non-Abelian ``magnetic" fields. 
To this aim we have to consider atoms with more internal states (``flavors"). The gauge potentials that can 
be realized using standard manipulations, such as laser-assisted tunneling and lattice acceleration, can have practically 
arbitrary  matrix form in the space of ``flavors". In such non-Abelian potentials, the single-particle spectrum generally  
depicts a complex structure termed by one of us Hofstadter ``moth" \cite{Osterloh}, 
which is characterized by numerous extremely small gaps. The  model of Ref. \cite{Osterloh} 
has stimulated further investigations, 
including  studies of  nontrivial quantum transport properties \cite{Clark}, as well as studies of 
the integer quantum Hall effect (IQHE) for cold atoms \cite{Goldman1}, spatial patterns in optical lattices 
\cite{Goldman2}, modifications of the Landau levels \cite{santos}, and quantum atom optics \cite{santos2,santos3}.

One should note,
however, that  the $U(2)$ gauge potentials proposed in Ref. \cite{Osterloh} and used in most of the following works are 
 characterized by \emph{non-constant} Wilson loops: atoms performing a loop around a plaquette undergo a unitary transformation which depends linearly on one of the spatial coordinates. Although such gauge potentials are interesting {\it per se}, the features characterizing the 
Hofstadter ``moth"  result from this {\it linear spatial
 dependence} of the Wilson loop, rather than from their non-Abelian nature.
 Indeed, the Hofstadter ``moth"-like spectrum  may actually be found in the standard Abelian case 
with a Wilson loop proportional to  $x$ (see Fig. 1). 

Two of us have shown that cold fermionic atoms trapped in optical lattices and 
subjected to artificial ``magnetic" fields should exhibit an IQHE \cite{Goldman1}. If a static force 
is applied to atoms, for instance  by accelerating the lattice, the
transverse Hall conductivity gives the
relation between this external forcing and the transverse atomic
current through the lattice. It has been shown that this transverse conductivity is quantized, 
$\sigma _{xy}=-\frac{C}{h}$, where $C$ is an integer 
and $h$ is Planck's constant.  Note that this quantity can be easily measured
from density profiles, as shown recently by Umucalilar {\it et al.} \cite{Umu}.
 The quantization
of $\sigma _{xy}$ occurs, however, only if  the Fermi energy of the system is located  inside a gap of the single-particle spectrum. While the observation of the IQHE  
seems to be  experimentally feasible in Abelian ``magnetic" fields, it is hardly so in the deeply non-Abelian regime in which 
the gaps of the ``moth" become very small \cite{rem}. 

The question therefore arises whether the consideration of non-Abelian gauge potentials characterized by a \emph{constant} Wilson loop could stabilize the spectral gaps and guarantee the robustness of the IQHE in ultracold fermionic gases
and whether an anomalous IQHE, as observed in graphene, can exist in such systems.

\begin{center} 
\begin{figure}
{\scalebox{0.24}{\includegraphics{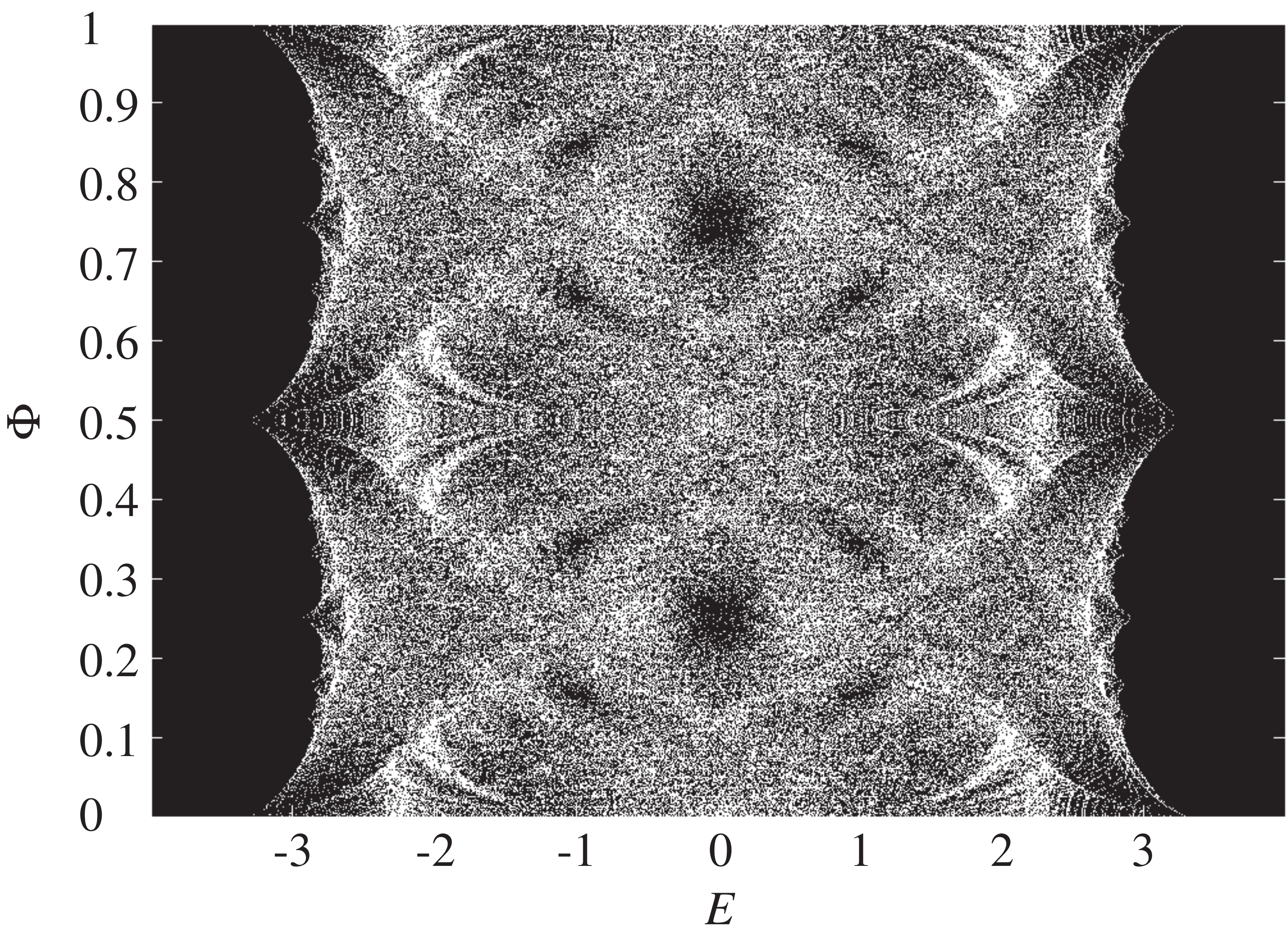}}} 
\caption{\label{abel} Energy spectrum $E=E(\Phi)$ in the case of the Abelian gauge potential $ \boldsymbol{A}=(0, 2 \pi \Phi m^2,0)$ with $x=ma$, corresponding to 
the \emph{non-constant} Wilson loop $W(m)=e^{i 2 \pi \Phi (1+2 m)}$. Compare with the Hofstadter ``moth" depicted in Fig. 1 of Ref. \cite{Goldman1} or in Fig. 4 of Ref. \cite{Travel}.} 
\end{figure} 
\end{center} 

In this work we provide affirmative answers to these questions by considering the IQHE in a system that features a  non-Abelian gauge potential characterized by  specific  non-commutating constant components and by a \emph{constant} Wilson loop. 
We calculate the energy spectrum and we obtain a robust
band structure with well developed gaps, which differs drastically from the case of the gauge potential of Ref. \cite{Osterloh}. In particular, we note the existence of van Hove singularities in the density of states and we obtain their analytical expression. We then evaluate the conductivity
$\sigma_{xy}$ for neutral currents using topological methods: we express $\sigma_{xy}$  in terms of the topologically invariant Chern numbers associated to each energy band \cite{remark}.  We eventually
present a salient fractal phase diagram which represents 
the integer values of the transverse conductivity inside the infinitely many
gaps of the spectrum. In this way, we show that the IQHE survives in  the non-Abelian regime, but undergoes strong modifications 
with striking similarity to the anomalous IQHE in graphene \cite{graphene}: the transverse conductivity suddenly changes sign due to the presence of van Hove singularities and is, under certain conditions, anomalous because of conical energy spectra.

\section{Optical lattice coupled to a non-Abelian gauge potential}
We consider a system of non-interacting two-component fermionic atoms trapped in a 2D optical square lattice of unit 
length $a$, with sites at $(x=m a,y=n a)$, with $n,m$ integers. The non-interacting limit can be reached using Feshbach resonances, or simply at low densities. The optical potential is strong, so that  the tight-binding approximation holds. The Schr\"odinger equation for a single particle subjected to an artificial gauge potential then reads 
\begin{align}
&t_a (U_x \, \psi_{m+1,n}+  U_x^{\dagger} \, \psi_{m-1,n}) \notag \\
&+  t_b (U_y \, \psi_{m,n+1}+  U_y^{\dagger} \, \psi_{m,n-1})=E \, \psi_{m,n} ,
\label{ham}
\end{align}
where $U_x$ (resp. $U_y$) is the tunneling operator and $t_a$ (resp. $t_b$) is the tunneling amplitude in the $x$ (resp. $y$) direction. In the following, we use $a$ as the length, and $t_a=t_b=t$ as the energy units, and set $\hbar=c=e=1$, except otherwise stated. The tunneling operators are related to the gauge potential according to $U_x=e^{i A_x}$.

Here we consider a general non-Abelian gauge potential
\begin{equation}
\boldsymbol{A}= \bigl ( \alpha  \sigma_y , 2 \pi \Phi m +\beta \sigma_x , 0 \bigr ) ,
\label{gauge}
\end{equation}
where $\alpha$ and $\beta$ are parameters, $(\sigma_x$, $\sigma_y)$ are Pauli matrices
 and $\Phi$ is the number of Abelian magnetic flux quanta per unit cell. 
 
In order to realize such a potential we may consider the method of  Ref. \cite{Osterloh}. However, the specific form of this gauge potential allows us to consider an even more practical scheme based on a  generalization of the method  currently developed by Klein and Jaksch \cite{imme}. We may use $^{40}$K  atoms in  $F=9/2$ or $F=7/2$ hyperfine manifolds, or $^6$Li with $F=1/2$. For $^{40}$K  one should optically pump and restrict the atomic dynamics to the two lowest Zeeman sublevels in each of the 
hyperfine manifolds. One can then employ different lattice tiltings in the $x$ and $y$ directions to perform laser (Raman assisted) tunnelings that change the internal states of the atoms; this allows to control the parameters $\alpha$ and $\beta$, and fixes the tunneling rate. Finally, the immersion of the system in a rotating BEC  will allow to control $\Phi$ \cite{imme}.  In experiments, one routinely reaches the  values of (laser assisted, or direct) tunneling 
rates in the range of 5-10 kHz ($\simeq$ 0.5 $\mu$K), Fermi temperatures  of the same order, and temperatures $T\simeq$ 0.2 $T_{\rm F}\simeq$ 50-100 nK (see for instance \cite{ouradv,Jaksch}). \\
The tunneling operators are $2 \times 2$ unitary matrices, 
\begin{align}
&U_x=\cos \alpha+i \sigma_y \sin \alpha , \notag \\
&U_y(m)=e^{i 2 \pi \Phi m} (\cos \beta +i \sigma_x \sin \beta ) ,
 \end{align}
 which act on the two-component wave function $\psi_{m,n}$. 

The single-particle Hamiltonian is invariant under translations defined by the 
operators $T^{q}_x \, \psi_{m,n}=\psi_{m+q,n}$ and $T_y \,  \psi_{m,n}=\psi_{m,n+1}$ under 
the condition that  $\Phi = \frac{p}{q}$, where $p$ and $q$ are integers.  Consequently, the system is 
restricted to a $q \times 1$ super-cell and one can express the wave function as 
$ \psi_{m,n}=e^{i k_x m} e^{i k_y n} u_m$ , with $u_m$ a $q$-periodic function. The wave 
vector $\boldsymbol{k}$ belongs to the first Brillouin zone, a $2$-torus defined as 
$k_x \in [0, \frac{2 \pi}{q}]$ and  $k_y \in [0,2 \pi]$. The Schr\"odinger equation \eqref{ham}  then reduces 
 to a generalized Harper equation
\begin{widetext}
\begin{align}
E \, u_m=& \begin{pmatrix}
\cos \alpha  & \sin \alpha \\ -\sin \alpha & \cos \alpha  \end{pmatrix} u_{m+1} e^{i k_x}   +  \begin{pmatrix}
\cos \alpha  & -\sin \alpha \\ \sin \alpha & \cos \alpha  \end{pmatrix} u_{m-1} e^{-i k_x}   \notag \\
&+ 2 \begin{pmatrix}
\cos (2 \pi \Phi m + k_y) \cos \beta   &  -\sin(2 \pi \Phi m + k_y) \sin \beta \\ - \sin (2 \pi \Phi m + k_y) \sin \beta &  \cos (2 \pi \Phi m + k_y) \cos \beta  \end{pmatrix} u_m .
\label{harper}
\end{align}
\end{widetext}

\section{The Non-Abelian regime}

\begin{center} 
\begin{figure}
{\scalebox{0.34}{\includegraphics{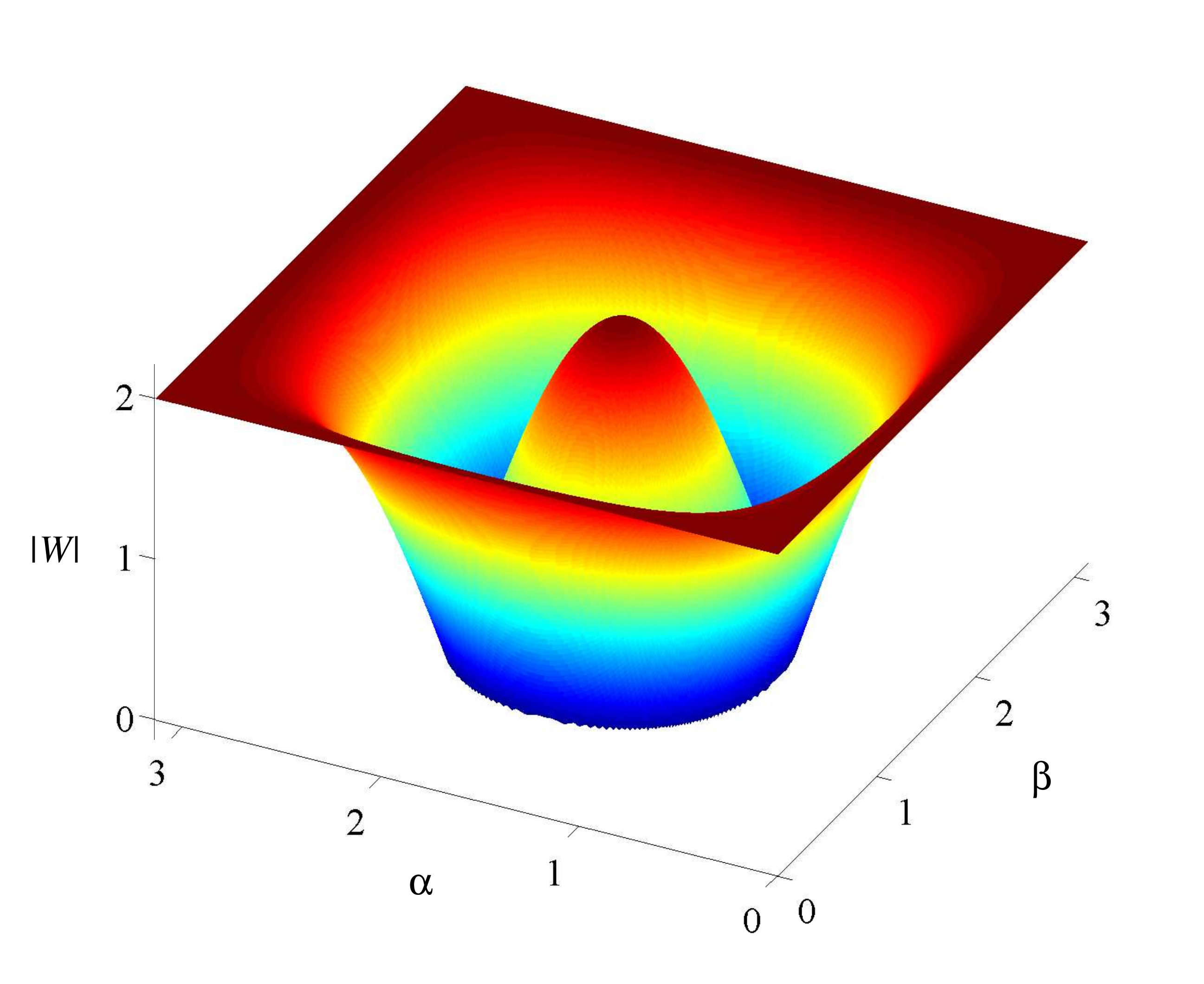}}} 
\caption{\label{wilson} Wilson loop's magnitude as a function of the parameters $\vert W \vert= \vert W (\alpha , \beta) \vert$. The system is equivalent to the Abelian Hofstadter model along the lines $\alpha=d \pi$ or  $\beta=d \pi$, where $d \in \mathbb{Z}$ and is equivalent to the Abelian $\pi$-flux model  at singular points, $\alpha=\beta= (2 d +1) \pi /2$. For any other values of the parameters $\alpha$  and $\beta$, the system is non-Abelian.} 
\end{figure} 
\end{center}

Artificial gauge potentials generally induce the following non-trivial unitary transformation for atoms hopping around a plaquette of the lattice:
\be
U=U_x U_y (m+1) U_x^{\dagger} U_y^{\dagger} (m).
\ee
 In the presence of the gauge potential  Eq. \eqref{gauge}, atoms performing a loop around a plaquette undergo the unitary transformation:
\begin{widetext}
\be
U= e^{i 2 \pi \Phi} \begin{pmatrix}
\cos ^2 \alpha + \cos 2 \beta \sin ^2 \alpha + \frac{i}{2} \sin 2 \alpha \sin 2 \beta 
& \sin 2\alpha \sin^2 \beta - i \sin^2 \alpha \sin 2\beta \\
-\sin 2\alpha \sin^2 \beta - i \sin^2 \alpha \sin 2\beta 
& \cos ^2 \alpha + \cos 2 \beta \sin ^2 \alpha - \frac{i}{2} \sin 2\alpha \sin 2\beta
\end{pmatrix}.
\ee
\end{widetext}
If one sets $\alpha=d \pi$ or  $\beta=d \pi$, where $d$ is an integer, the matrix  $U= \exp (i 2 \pi \Phi)$ is proportional to the identity and the system behaves similarly to the Hofstadter model \cite{Hofstadter}. When $\alpha=\beta= (2 d +1) \pi /2$, where $d \in \mathbb{Z}$, one finds that $U= -\exp (i 2 \pi \Phi)$ and the system is equivalent to the $\pi$-flux model in which half a flux quanta is added in each plaquette \cite{graphene}. In these particular cases where $U= \pm \, e^{i 2 \pi \Phi}$, the system is in the Abelian regime. For any other values of the parameters $\alpha$ and $\beta$, the matrix $U$ is a non-trivial $U(2)$ matrix and the system is non-Abelian. \\ 
A gauge invariant quantity which characterizes the system is given by the Wilson loop
\begin{align}
&W=\textrm{tr} \, U_x U_y (m+1) U_x^{\dagger} U_y^{\dagger} (m) \notag \\
&= 2 \, e^{i 2 \pi \Phi} \, (\cos ^2 \alpha + \cos 2 \beta \sin ^2 \alpha )
\end{align}
It is straightforward to verify that the system is non-Abelian when $\vert W \vert \ne 2$, and that $W (\alpha , \beta)=W (\beta , \alpha )$. In Fig. \ref{wilson}, where we show the Wilson loop's magnitude as a function of the parameters, $\vert W \vert= \vert W (\alpha , \beta) \vert$,  we can easily identify the regions corresponding to the Abelian ($\vert W \vert = 2$) and to the non-Abelian regimes ($\vert W \vert \ne 2$). We note that the Abelian $\pi$-flux regime is reached at a singular point, $\alpha=\beta= (2 d +1) \pi /2$, where $d \in \mathbb{Z}$.  

We also point out that the statement
according to which the non-Abelian regime is reached when $[U_x,U_y] \ne 0$, and which can be found in previous works \cite{Osterloh,Goldman1}, is incorrect: for the situation where $\alpha=\beta= (2 d +1) \pi /2$, one finds that $[U_x,U_y] =2 i \, e^{2 i m \pi \Phi} \sigma_z$, while the system is Abelian because of its trivial Wilson loop, $\vert W \vert=\vert - 2 e^{2 i \pi \Phi} \vert =2$.

Contrary to the non-Abelian systems considered in previous works \cite{Osterloh,Clark,Goldman1}, we emphasize that the gauge potential Eq. \eqref{gauge} leads to a Wilson loop which does not depend on the spatial coordinates. In the following section, we show that this feature leads to energy spectra and fractal structures which significantly differ from the Hofstadter ``moth" \cite{Osterloh}.

\section{The energy spectrum}

The energy spectrum can be obtained through direct diagonalization of  Eq. \eqref{harper}. 

In the Abelian regime 
corresponding to $\alpha = d \pi$ or $\beta = d \pi$, where $d \in \mathbb{Z}$, one finds $q$ doubly-degenerated bands 
for $\Phi=\frac{p}{q}$. In this particular case, the representation of the spectrum  as a function of the flux $\Phi$ 
leads to the fractal Hofstadter ``butterfly" \cite{Hofstadter}. For the other Abelian case $\alpha=\beta=\frac{\pi}{2}$, the system behaves according to the $\pi$-flux lattice: the spectrum $E=E(\Phi)$ depicts a Hofstadter ``butterfly" which is contained between $\Phi=[0.5 ; 1.5]$, i. e. shifted by $\Phi=0.5$ with respect to the original ``butterfly", and the system remarkably describes zero-mass Dirac particles  \cite{graphene}.  

\begin{center} 
\begin{figure}
\begin{center} 
{\scalebox{0.28}{\includegraphics{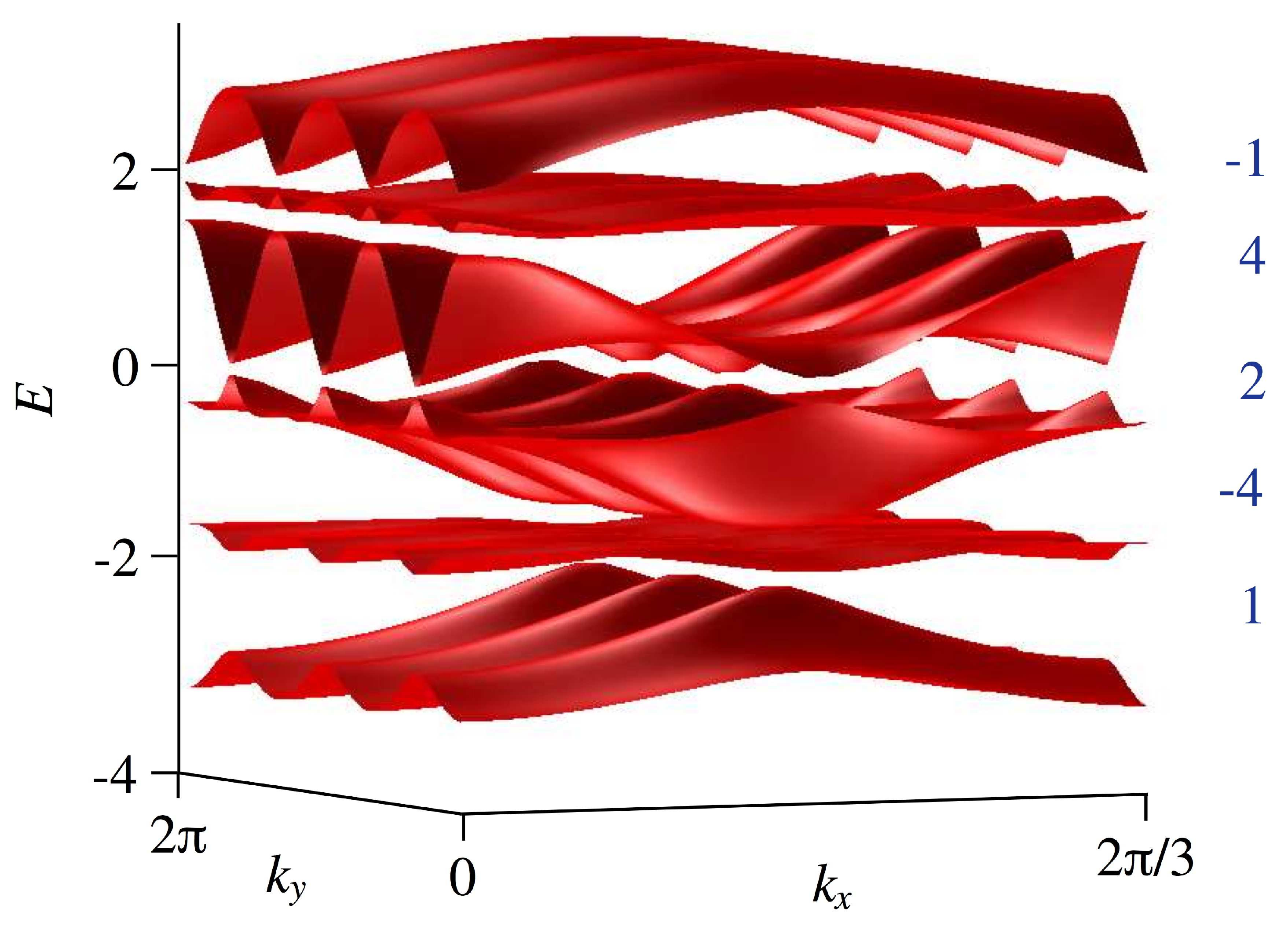}}} 
\caption{\label{bandfig} (Color online) Spectrum $E=E (k_x, k_y)$ for $\alpha=\beta=1$ and $\Phi=\frac{1}{3}$. When the Fermi energy lies within a gap, the transverse conductivity $h \, \sigma_{x y}$ is quantized (blue integers).} 
\end{center}
\end{figure} 
\end{center}

\begin{center} 
\begin{figure}
\begin{center} 
{\scalebox{0.21}{\includegraphics{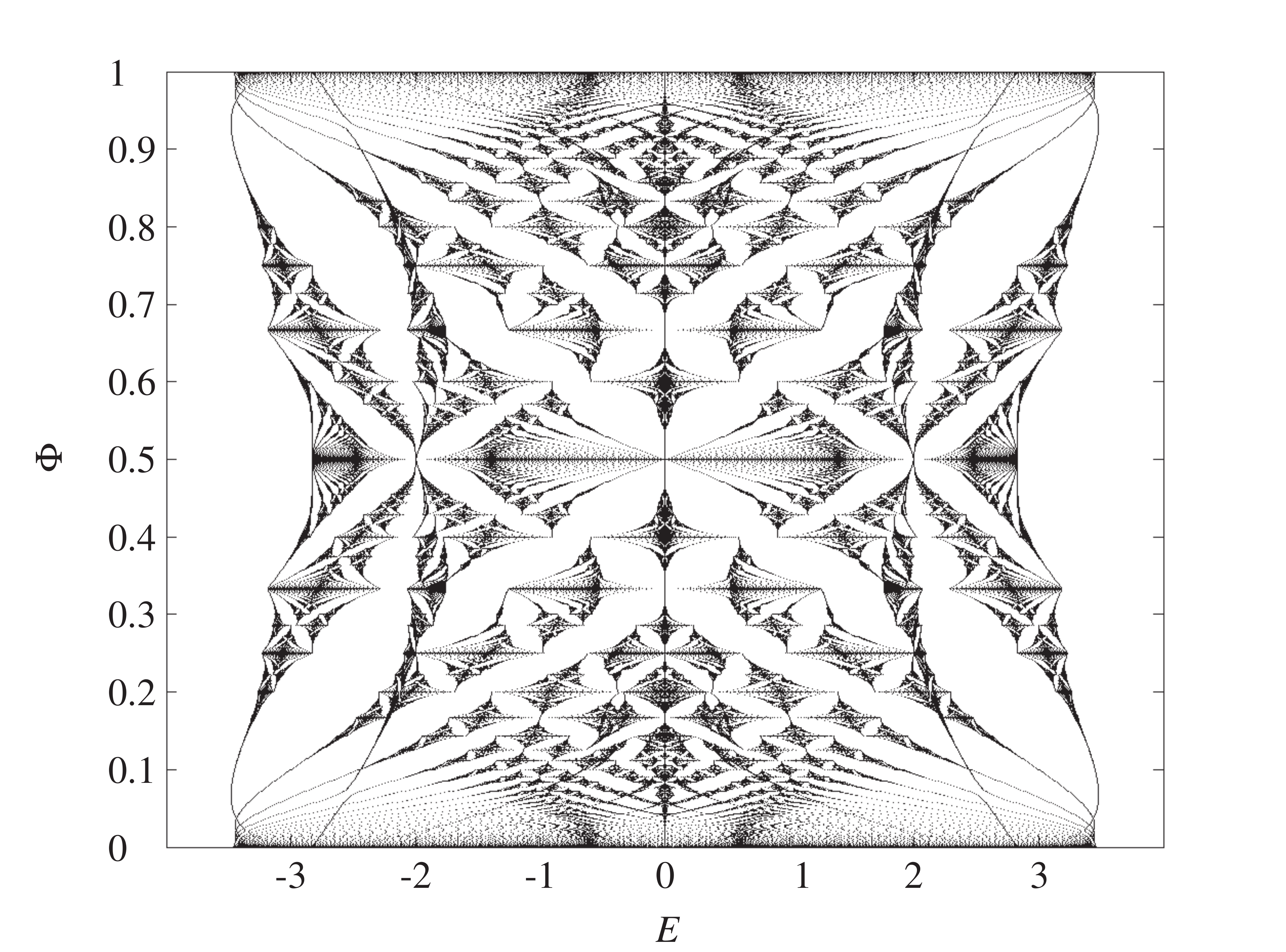}}} 
\caption{\label{flux} Spectrum $E=E (\Phi)$ for $\alpha=\beta=\frac{\pi}{4}$ and $\Phi=\frac{p}{827}$, where $p$ is an integer. When $\Phi=0$, four van Hove singularities are located at $E= \pm (2 + \sqrt{2})$ and $E= \pm (2 - \sqrt{2})$.} 
\end{center}
\end{figure} 
\end{center}

In the non-Abelian regime, which is 
reached for arbitrary values of the parameters $(\alpha, \beta)$, the spectrum is constituted of $2 q$ separated bands as 
illustrated in Fig. \ref{bandfig}. For these general situations, the representation of the spectrum as a function of the flux $\Phi$  leads to new interesting features. As in the Abelian case, one observes repetitions of similar structures at various scales. However, new patterns arise in the non-Abelian case, as illustrated in Fig. \ref{flux} for $\alpha=\beta=\frac{\pi}{4}$ and in Fig. \ref{one} for $\alpha=1$ and $\beta=2$. It is worth noticing that for arbitrary values of the 
 parameters $(\alpha, \beta)$, the spectra show well-developed gaps contrasting with the Hofstadter ``moth" which appears in the non-Abelian system proposed in Ref. \cite{Osterloh}.  We further notice that the spectrum is periodic with period $T_{\Phi}=1$ and is symmetric with respect to  $E=0$ and $\Phi=0.5$. 
 
 In the non-Abelian regime close to $\alpha, \beta =\pi /2$, one observes that  conical intersections are preserved in the energy spectrum. As shown in the next section, the particles behave similarly to Dirac particles in this non-Abelian region and the system exhibits an anomalous quantum Hall effect.
 
 We eventually note that when the flux $\Phi=0$, the density of states reveals several van Hove singularities at the energies $E= \pm 2 (1 + \cos \chi)$ and $E= \pm 2 (1 - \cos \chi)$, where $\chi=\alpha , \beta$. As the flux increases, these singularities evolve and generally merge. 
 
\section{Integer quantum Hall effect and the phase diagram}

We evaluate the linear response of the system described by Eq. \eqref{harper} to
an external force (lattice acceleration) applied along the $y$ direction and we evaluate the 
transverse conductivity $\sigma_{x y}$ using  Kubo's formula. Following the method of Ref. \cite{Goldman1}, one can generalize the well-known TKNN expression \cite{Thoules} to the present non-Abelian framework, yielding
\begin{align}
\sigma_{x y}&= \frac {1}{2 \pi  i h}  \sum_{E _{\lambda}
< E _{\rm F}}  \int_{\mathbb{T}^2} \sum_j \biggl ( \langle
\partial _{k_{x}} u_{\lambda j} \vert \partial _{k_{y}} u_{ \lambda
j} \rangle  \notag \\ 
&\qquad\qquad\qquad\qquad -\langle
\partial _{k_{y}} u_{\lambda j} \vert \partial _{k_{x}} u_{ \lambda
j} \rangle  \biggr ) d\boldsymbol{k} ,
\label{hall}
\end{align}  
where $u_{\lambda j}$ is the $j$th component of the wave function corresponding to the band $E_{\lambda}$ such that $H u_{\lambda}= E_{\lambda} u_{\lambda}$, and $\mathbb{T}^2$ refers to the first Brillouin zone of the system. The Fermi energy $E _{\rm F}$ is supposed to lie within a gap of the spectrum. The transverse conductivity  is then given by the contribution of all the states filling the bands $E _{\lambda}< E _{\rm F}$ situated below this gap. 

Eq. \eqref{hall} conceals a profound topological interpretation for the transverse conductivity  based on the fibre bundle theory \cite{Nakahara}. In the present framework, such bundles are conceived as the product of the parameter space $\mathbb{T}^2$ with the non-Abelian gauge group $U(2)$. This product space, which is supposed to be locally trivial but is generally expected to twist globally, is characterized by the non-Abelian Berry's curvature 
\be
\mathcal{F}=\bigl ( \partial_{k_x} \mathcal{A}^{y}-\partial_{k_y} \mathcal{A}^{x} + [\mathcal{A}^{x}, \mathcal{A}^{y}] \bigr ) dk_{x} dk_{y},
\ee
 where $(\mathcal{A}^{\mu})_{i j}=\langle u_{\lambda i} \vert  \partial _{k_{\mu}} u_{\lambda j} \rangle $ is the Berry's connection. The triviality of the fibre bundle is measured by the Chern number 
 \be
 C(E_{\lambda})= \frac{i}{2 \pi} \int_{\mathbb{T}^2} {\rm tr} \mathcal{F} ,
 \ee 
 which is a topological invariant and is necessarily an integer. Note that each band $E_{\lambda}$ is associated to a specific fibre bundle, on which a Chern number is defined. One eventually finds that the Hall-like conductivity Eq. \eqref{hall} is given by a sum of integer Chern numbers, 
\be
\sigma_{x y}= -\frac {1}{h}  \sum_{E _{\lambda}
< E _{\rm F}}  C(E_{\lambda}).
\label{final}
\ee
 As a consequence, the transverse Hall-like 
conductivity of the system evolves by steps corresponding to integer multiples of the 
inverse of Planck's constant and is robust against small perturbations.

The evaluation of these topological invariants leads to a complete understanding
of the IQHE which takes place in the present context. The aim is then to compute the 
Chern number associated to each band $E _{\lambda}$ of the spectrum. This computation can be achieved numerically thanks to an efficient method developed by  Fukui {\it et al.} \cite{Fukui} and which can be applied to our specific system. This method is summarized as follows: the Brillouin zone $\mathbb{T}^2$, defined by $k_x \in [0, \frac{2 \pi}{q}]$ and  $k_y \in [0, 2 \pi]$, is discretized into a lattice constituted by points denoted $\boldsymbol{k}_l=(k_{xl},k_{yl})$. On the lattice one defines a curvature $\mathcal{F}$ expressed as 
\be
\mathcal{F}_{12} (\boldsymbol{k}_l)= \textrm{ln} \, U_1 (\boldsymbol{k}_l) U_2 (\boldsymbol{k}_l +\hat{ \boldsymbol{1}}) U_1 (\boldsymbol{k}_l +\hat{ \boldsymbol{2}})^{-1} U_2 (\boldsymbol{k}_l)^{-1} , 
\ee
where the principal branch of the logarithm with $- \pi <  \mathcal{F}_{12}/i  \le \pi$ is taken,
$\hat{\boldsymbol{\mu}}$ is a unit vector in the direction $\mu$, and
\be
U_{\mu} (\boldsymbol{k}_l)= \sum_j \langle u_{\lambda j} (\boldsymbol{k}_l) \vert u_{\lambda j} (\boldsymbol{k}_l+\hat{ \boldsymbol{\mu}}) \rangle / \mathcal{N}_{\mu} (\boldsymbol{k}_l),
\ee
defines a link variable with a normalization factor $\mathcal{N}_{\mu} (\boldsymbol{k}_l)$ such that $\vert U_{\mu} (\boldsymbol{k}_l)\vert=1$. The Chern number associated to the band $E_{\lambda}$ is then defined by 
\be 
C=\frac{1}{2 \pi i} \sum_l \mathcal{F}_{12} (\boldsymbol{k}_l).
\ee
 This method ensures the integral character of the Chern numbers and holds for non-overlapping bands. In the situations where the spectrum reveals band crossings, a more general definition of the link variables $U_{\mu} (\boldsymbol{k}_l)$ has been proposed in Ref. \cite{Fukui}. 
 
 We first compute the Chern numbers for a specific case, illustrated in Fig. \ref{bandfig}. For $\alpha=\beta=1$ and $\Phi=\frac{1}{3}$, the Chern numbers associated to the six bands are respectively ${ 1 ; -5 ; 6 ; 2 ; -5 ; 1}$. According to Eq. \eqref{final}, the transverse conductivity's values associated to the 5 gaps are ${ 1 ; -4 ; 2 ; 4 ; -1}$ as shown in Fig. \ref{bandfig}.

\begin{center} 
\begin{figure}[h!]
{\scalebox{1.91}{\includegraphics{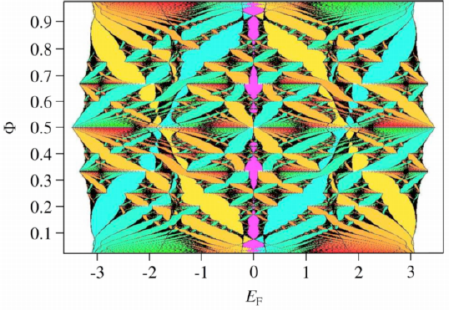}}} 
\caption{\label{one} (Color online) Spectrum $E=E(\Phi)$ and phase diagram for $\alpha =1, \beta = 2$ and $\Phi =\frac{p}{q}$ with $q<97$. 
Warm (resp. cold) colors correspond to positive (resp. negative) values of the quantized conductivity. 
Purple corresponds to a null transverse conductivity. For $\Phi \ll 1$, the quantized conductivity evolves monotonically but suddenly changes sign around the van Hove singularities located at $E \simeq \pm 1$ (see the alternation of cold and warm colors).} 
\end{figure} 
\end{center}

The phase diagram describing the IQHE for our model can eventually be drawn. 
 In this diagram we represent  the quantized transverse conductivity as a function of the Fermi 
energy $E _{\rm F}$ and flux $\Phi$. Here we illustrate a representative example of such a phase diagram 
which was obtained for $\alpha =1$, $\beta = 2$ (cf. Fig. \ref{one}). This striking figure differs radically 
from the phase diagrams obtained by Osadchy and Avron in the Abelian case \cite{Osadchy} since the Chern numbers associated to the gaps are no 
longer satisfying a simple Diophantine equation \cite{dioremark}. Consequently, the measurement of the transverse conductivity 
in this system should show a specific sequence of robust plateaus,  
heralding a new type of quantum Hall effect. 

This new effect is comparable to 
the IQHE observed in Si-MOSFET or  the anomalous IQHE observed in graphene
in the ``low flux" regime $\Phi \ll 1$ corresponding to experimentally available magnetic fields.
In this regime, the quantized conductivity evolves monotonically by steps of one between sudden changes of sign across the aforementioned van Hove singularities (see Fig. \ref{one}). Moreover, in the vicinity of $\alpha , \beta = \pi/2$, the quantized conductivity increases by double integers because of Dirac points in the energy spectrum, in close similarity with the anomalous IQHE observed in graphene.

\section{Conclusions}

Summarizing, we have proposed in this paper how to realize in cold atomic systems a textbook example of 
 non-Abelian gauge potential characterized by a \emph{constant} Wilson loop. 
Our main result is that despite the coupling between the different ``flavor" components of the 
single-particle wave functions, the spectrum exhibits well-developed gaps of order of 0.1-0.2$t$, 
i.e. about 50-100 nK. 

The IQHE survives in the deeply non-Abelian regime and acquires a unique character 
specific to the non-Abelian nature of the gauge fields. It is characterized by a particular sequence of 
robust plateaus corresponding to the quantized values of the transverse conductivity. Moreover, the 
non-Abelian coupling induces controllable van Hove singularities as well as an anomalous Hall effect, similar to the effect induced by the hexagonal geometry in graphene. Experimental observation of this distinctive effect requires to achieve $T$ smaller than the gaps, 
i.e. of order of 10-50 nK, which is demanding but not impossible. 

The main experimental challenge 
consists here in combining several established methods into one experiment: 
laser assisted tunneling \cite{Jaksch}, BEC immersion \cite{imme}, and density profile measurements \cite{Umu}.

We acknowledge support of the EU IP Programme SCALA, ESF-MEC Euroquam Project FerMix, Spanish MEC grants (FIS 2005-04627, 
Conslider Ingenio 2010 ``QOIT), the Belgian Federal Government (IAP project ``NOSY"), and the "Communaut\'e fran\c caise de Belgique" (contract "ARC" No. 04/09-312) and the F.R.S.-FNRS Belgium. M.L acknowledges ERC AdG "QUAGATUA". N. G. thanks S. Goldman for his comments, Pierre de Buyl for its support and ICFO for its hospitality. A.K. acknowledges support of the Polish Government Research Grant for 2006-2009 and thanks J. Korbicz and C. Menotti for discussions. The authors thank D. Jaksch, J. V. Porto, and C. Salomon for their valuable insights in various aspects of this work.



\begin{thebibliography}{99}
\bibitem{rem}{Close to the Abelian regime, the Hall conductivity can be explicitly computed as some large gaps persist inside the ``moth" spectrum \cite{Goldman1}.}
\bibitem{remark}{Chern numbers \cite{Nakahara} have been used in the IQHE context \cite{Thoules}, as well as to characterize strongly correlated states \cite{Wen1,Hafezi1}. The Chern analysis was performed in Refs. \cite{Goldman1,Goldman3} in order to characterize the IQHE in atomic systems and quantum graphs.}
\bibitem{dioremark} The quantized conductivity depends on the real-valued parameters $\alpha$ and $\beta$. Thus a Diophantine equation cannot be established here, contrary to the Abelian case \cite{Thoules}.
\end{thebibliography}
\end{document}